\documentclass[11pt,a4paper]{article}
\usepackage{amssymb}
\usepackage{amsmath}
\usepackage[pdftex]{graphicx}
\usepackage{graphics}
\usepackage{epsfig}
\usepackage{color}
 
\newcommand{\be}{\begin{equation}}
\newcommand{\ee}{\end{equation}}
\numberwithin{equation}{section}

\allowdisplaybreaks

\begin{document}

\title{\bf The Birman-Schwinger operator for the Cornell Hamiltonian}

 \author{O Civitarese$^1$, S. Fassari$^2$, M. Gadella$^3$, F. Rinaldi$^2$\\ \\
$^1$Department of Physics. National University of La Plata, \\ 1990 La Plata, Argentina\\  $^2$Dipartimento di Scienze Ingegneristiche, \\ Universit\`a Degli Studi Guglielmo Marconi, \\ Via Plinio 44,
I-00193 Rome, Italy \\
$^3$ Departamento de F\'{\i}sica Te\'orica, At\'omica y Optica  and IMUVA, \\
Universidad de Va\-lladolid, 47011 Valladolid, Spain.\\} 

\maketitle

\begin{abstract}
{Quantum Chromodynamics is the theory of strong interactions. It has been shown during the last decades that it describes correctly most of the properties of hadrons at high energy. The most distinctive feature of the theory is the realisation that the elementary particles which composed the known forms of matter, that is to say quarks and gluons, cannot be observed at low energy. In this work we are addressing this specific feature, known as confinement, by performing  a rigorous mathematical treatment of the Cornell potential.}
\end{abstract}

\section{Introduction}\label{sec1}

The term Quantum Chromodynamics (QCD) refers to the theory of strong interactions successfully developed over the last 60 years \cite{weinberg}. It describes the phenomenology of particle physics at high energy, which is the domain where perturbative expansions can be applied to describe observables \cite{tdlee}. Based on fermionic (quarks) and bosonic (gluons) degrees of freedom and their interactions, QCD has passed quite a few crucial tests.
However, at low energies, the theory becomes non-perturbative and various attempts to overcome this difficulty have been proposed in the past, it may be worth recalling among them the Lattice Gauge Theory \cite{lgt1, lgt2}, the Dyson-Schwinger expansion method \cite{ds1, ds2}, as well as methods based on Group Theory (\cite{so40}-\cite{so43}).

The main feature of QCD at low energy, that is to say the impossibility of observing free quarks, a property known as confinement, was emphasized in the early eighties by Godfrey and Isgur in their pioneering paper \cite{gi}. It was later expressed formally by means of the so-called Cornell Potential \cite{weinberg} modelling the competing effects between the Coulomb interaction and the linear one taking place between quarks, where the most distinctive feature is the running value of the confinement coupling $\alpha_s(q)$ as a function of the momentum. 

The literature is very rich concerning the theoretical and experimental implications of the use of the Cornell Potential to describe observed features, for instance, of baryonic and mesonic systems \cite{data}. Its use in effective theories of QCD at low energy ({\cite{heff1}-\cite{heff4}}) provided a good evidence about its capability to account for the hadronic spectra at low energy.

In this article we shall focus our attention on the mathematical aspects of the problem in order to explore the validity of some of the approximations which are usually applied in dealing with the Cornell Potential.
To achieve this goal we have developed an application of Birman-Schwinger operator method and extracted from it expressions for the eigenvalues of a Coulomb plus linear potential.

The details of the derivation are presented in the next sections, as well as the conclusions drawn from our study. The paper is organised as follows: Section 2 describes the structure of the Hamiltonian with the Cornell Potential. By regarding the Coulomb part of the interaction as a perturbation, the Birman-Schwinger (BS) method is applied. The detailed analysis of the BS integral operator is carried out in Section 3. In Section 4 the wave functions obtained by means of the BS method are given. Finally, the conclusions are presented in Section 5.

\section{The Hamiltonian}\label{sec2}

The objective of the present contribution is the study of the bound states of the Hamiltonian

\begin{equation}\label{1}
H= -\frac{\hbar^2}{2m}\, \frac{d^2}{dr^2} +V_l r - \frac{V_C}{r} = H_0 - \frac{V_C}{r} \,, \qquad r>0\,,
\end{equation}
where $V_l$ and $V_C$ are positive constants. This is a one-dimensional version of the Cornell potential described in \cite{gi,cornell2}.

In the study of the Hamiltonian \eqref{1}, it is convenient to start with the first summand $H_0$, regarded as the
unperturbed Hamiltonian:

\begin{equation}\label{2}
H_0 = -\frac{\hbar^2 c^2}{2mc^2}\, \frac{d^2}{dr^2} +V_l r \, \qquad r>0 \,.
\end{equation}

It may be worth pointing out that the one-dimensional Hamiltonian on the entire real line with the confining potential $V(x)=|x|$ perturbed by an attractive point interaction, given by a Dirac $\delta$ or a local $\delta'$ or a nonlocal $\delta'$, has been thoroughly studied in \cite{AnnPhys17}. Furthermore, its two-dimensional analogue with the confining potential $V(x,y)=|x|+|y|$ perturbed by an attractive 2D Dirac $\delta$ has been studied in \cite{FGGNR}.

The Schr\"odinger equation associated with $H_0$ reads

\begin{equation}\label{3}
- \frac{d^2}{dr^2}\, \psi(r) + \frac{2mc^2}{\hbar^2 c^2} \, V_l \, r \psi(r) - \frac{2mc^2}{\hbar^2 c^2} \, E \, \psi(r) =0\,.
\end{equation}

With the change of variables,

\begin{equation}\label{4}
z:= \lambda_0 r - \beta\,, \qquad \lambda_0 = \frac{2mc^2 V_l}{\hbar^2 c^2} \,, \qquad \beta = \frac{2mc^2 E}{\hbar^2 c^2}
\end{equation}
equation \eqref{3} takes the following form:

\begin{equation}\label{5}
\varphi''(z) - \omega^2 \, z \, \varphi(z)\,=0, \qquad \omega^2 = \frac{1}{\lambda_0^2} \,, \qquad \varphi(z) := \psi\left(\frac{\beta + z}{\lambda_0} \right)\,,
\end{equation}
and the prime represents derivative with respect to the variable $z$.

Equation \eqref{5} is quite similar to the Airy differential equation. In order to arrive at this equation, we perform the following change of scale:

\begin{equation}\label{6}
\eta:= \frac{z}{\gamma}\,,
\end{equation}
so that \eqref{5} takes the following form:

\begin{equation}\label{7}
\phi''(\eta) - \omega^2 \gamma^3 \eta\, \phi(\eta) =0\,,
\end{equation}
where the prime denotes derivative with respect to $\eta$. We fix the scale with the condition $\omega^2 \gamma^3 =1$, so that we finally arrive at the Airy differential equation:

\begin{equation}\label{8}
\phi''(\eta) - \eta \,\phi(\eta) =0\,, \qquad  r= \frac{\gamma\eta + \beta}{\lambda_0}\,.
\end{equation}

Here, since $r > 0$, we should impose a sort of Dirichlet boundary condition for the solutions of \eqref{8} of the
form:

\begin{equation}\label{9}
\lim_{r\to 0^+} \psi(r) = \psi(0^+) = \phi(-\beta/\gamma)\,.
\end{equation}

For the sake of simplicity, let us set $\alpha:= \beta/\gamma$. Then, the boundary condition \eqref{9} reads $\phi(-\alpha)=0$. Let us perform a final change of variables given by $v:= \eta + \alpha$ (recall that $\alpha>0$). Then, \eqref{8} becomes:

\begin{equation}\label{10}
\xi''(v) - (v - \alpha)\, \xi(v) =0\,, \qquad \xi(v) := \phi(v-\alpha)\,.
\end{equation}

Note that $\xi(0)= \phi(-\alpha)=0$. Then, 

\begin{equation}\label{11}
v= \eta + \alpha = \eta + \frac{\beta}{\gamma} = \frac{z + \beta}{\gamma} = \frac{\lambda_0}{\gamma}\, r\,,
\end{equation}
so that $r>0$ if and only if $v>0$. We have solutions with $r > 0$ only. As is well known (Section 8.1.1 in \cite{VS}), these solutions have
the following form:

\begin{equation}\label{12}
\xi_{n}(v) = C_{n} \, Ai(v- \alpha_{n})\,,
\end{equation}
where 

\noindent i) $Ai(-)$ is the Airy function and $-\alpha_{n}$ are the zeros of the $Ai(-)$ function, which, as attested by \cite{LM,AnnPhys17,VS}, are always located along $\left(-\infty,0\right]$, so that $Ai(v- \alpha_{n})$ vanishes at $v=0$; 

\noindent ii) As shown in \cite{LM,AnnPhys17,VS}, $ \alpha_{n}$ behaves like $n^{\frac 23}$ as $n\, \rightarrow \infty$; 

\noindent iii) $C_{n}$ are the required normalisation constants.

Due to the definition of $\alpha_n$ one has that

\begin{equation}\label{13}
\alpha_n= \left( \frac{2mc^2}{\hbar^2 c^2}\right)^{1/3} V_l^{-2/3} E_n \,,
\end{equation}
so that the energy levels are given by

\begin{equation}\label{14}
E_{n} = \left( \frac{\hbar^2 c^2}{2mc^2}  \right)^{1/3}\, V_l^{2/3}\, \alpha_{n} \,.
\end{equation}

Due to \eqref{12}, the corresponding eigenfunctions are given by

\begin{equation}\label{15}
\xi_{n}(v) = \frac{Ai(v- \alpha_{n})}{Ai'(-\alpha_{n})} = \frac{Ai(\frac{\lambda_0}{\gamma}\, r- \alpha_{n})}{Ai'(-\alpha_{n})} =  \psi_{n}(r)\,,
\end{equation}
for $n=0,1,2,\dots$. These functions form an orthonormal basis in $L^2(\mathbb R^+)$ (Section 8.1.1 in \cite{VS}). 

Finally, the expression of the Coulomb contribution of the potential in terms of the variable $\eta$ is given by

\begin{equation}\label{16}
\frac{V_C}{r} = \frac{\lambda_0 V_C}{\gamma \eta + \beta}\,,
\end{equation}
so that equation \eqref{1} takes the following form in terms of $\eta$:

\begin{equation}\label{17}
\phi''(\eta) - \eta\, \phi(\eta) - \frac{\lambda_0 V_C}{\gamma \eta + \beta} =0\,.
\end{equation}

Note that $\beta$ depends on the energy levels, so that each particular solution should take into account the particular value of $E_n$ obtained above.

\section{The Birman-Schwinger method}\label{sec3}

Once we have the solution of the spectral problem for $H_0$, it is possible to investigate the energy levels pertaining to \eqref{1} by regarding the Coulomb interaction as a perturbation provided a certain condition is met. To this end, we shall make use of the Birman-Schwinger method. The point of departure is the Schr\"odinger equation,

\begin{equation}\label{18}
(H_0 - V) \psi = E \psi \Longleftrightarrow (H_0 +|E|) \psi = V \psi \,,
\end{equation}
where $E<0$ and $V\geq 0$, which in this case is the Coulomb interaction $V_C/r$ with $V_C >0$ . Then, if $\chi:= V^{1/2} \psi$, we have that

\begin{equation}\label{19}
(H_0+|E|) V^{-1/2} V^{1/2} \psi = V^{1/2} V^{1/2} \, \psi \,,
\end{equation}
so that

\begin{equation}\label{20}
\chi= V^{1/2} (H_0 +| E|)^{-1} V^{1/2} \, \chi \,.
\end{equation}

The operator

\begin{equation}\label{21}
B_E:=V ^{1/2} (H_0 +|E|)^{-1} V^{1/2} \,,
\end{equation}
is the so-called {\it  Birman-Schwinger operator}. It is a positive integral operator and, due to the form of the potential $V$, its kernel is given by

\begin{equation}\label{22}
K(r,r';E) = \omega \sum_{n=0}^\infty  \frac{1}{r^{1/2}} \, \frac{Ai \left( \frac{\lambda_0}{\gamma}\, r -\alpha_{n} \right) \, Ai \left( \frac{\lambda_0}{\gamma}\, r' -\alpha_{n} \right)}{|Ai'(-\alpha_{n})|^2 \,(E_{n} + |E|)}\, \frac{1}{{r'}^{1/2}} \,, \quad \omega:= \frac{2mc^2\, V_c}{\hbar^2 c^2}\,.
\end{equation}

Note that the normalised eigenfunctions of $H_0$ are given by \eqref{15}. 

Before moving forward, we wish to point out that we have recently investigated the Birman-Schwinger operator pertaining to the one-dimensional Hamiltonian with the Coulomb potential without the linear confinement ($V_l=0$), showing that such a B-S operator belongs to the Hilbert-Schmidt class, namely the class of integral operators having a square summable  integral kernel (see \cite{FGLNR}).

The energy levels of \eqref{1} are those values of the energy corresponding to the lowest eigenvalue  of the Birman-Schwinger operator $B_E$ in \eqref{21}. In order to estimate those energy values by using some standard techniques, we  first  check whether $B_E$ is a trace class operator. To this end, by invoking the lemma used in Theorem XI.31 in \cite{RSIII}, we need only check whether the kernel $K(r,r;E)$ (with $r=r'$) is integrable in the variable $r$. Let us omit the parameter $\omega$ since it is clearly irrelevant in the following discussion. Then (note that $K(r,r;E)$ is non-negative),

\begin{eqnarray}\label{23}
\int_0^\infty K(r,r;E)\, dr = \int_0^\infty \frac 1r \sum_{n=0}^\infty \frac{Ai^2 \left( \frac{\lambda_0}{\gamma}\, r -\alpha_{n} \right)}{E_{n} + |E|}\, \frac 1{|Ai'(-\alpha_{n})|^2} \, dr. \nonumber \\ [2ex] = \sum_{n=0}^\infty \frac{1}{E_{n} + |E|}\, \frac 1{|Ai'(-\alpha_{n})|^2}. \int_0^\infty \frac 1r \, Ai^2 \left( \frac{\lambda_0}{\gamma}\, r -\alpha_{n} \right) \, dr \,.
\end{eqnarray}

The last integral in \eqref{23} always converges. Given that $-\alpha_{n}$ are the zeros of the Airy function, for $r = 0$ we always have that $Ai(-\alpha_{n}) =0$. Such functions are analytic at the origin, so that the last integral in \eqref{23} does not have divergences at the origin. In addition,  they decay rapidly at infinity since

\begin{equation}\label{24}
Ai^2(r) \approx \frac{1}{4\pi r^{1/2}}\, e^{-\frac 43  r^{3/2}}\,,
\end{equation}
for large values of $r$. Thus, we deduce the convergence of the aforementioned integral. The next step is to check whether the series in \eqref{23} converges. As a consequence of Corollary 3.6 in \cite{LM}, we have that

\begin{equation}\label{25}
\frac{1}{E_n +|E|} \le \left( \frac{8}{3\pi} \right)^{2/3} \, \frac1{(2n-1)^{2/3}} = \left( \frac{4}{3\pi} \right)^{2/3} \, \frac1{(n-1/2)^{2/3}} \,.
\end{equation}

According to A.9 in \cite{LM}, we have that

\begin{equation}\label{26}
|Ai'(-\alpha_n)|^2 \approx \frac 1\pi\, \left( \frac{3\pi}{2} \right)^{1/3} \left( n- \frac 12 \right)^{1/3}\,,
\end{equation}
so that,

\begin{equation}\label{27}
\frac 1{|Ai'(-\alpha_n)|^2} \approx \frac{2^{1/3} \pi}{(3\pi)^{1/3} (n-1/2)^{1/3}}\,.
\end{equation}

This shows that the product of the factors before the integral sign in the last term of \eqref{23} decays as rapidly as $(2n+1)^{-1}$. Then, in order to infer the convergence of the series it suffices to show that

This shows that the product of the factors before the integral sign in the last term of \eqref{23} decays as rapidly as $(2n+1)^{-1}$. Then, in order to infer the convergence of the series it suffices to show that

\begin{equation}\label{28}
\lim_{n\to \infty} \int_0^\infty \frac 1r \, Ai^2 \left( \frac{\lambda_0}{\gamma}\, r - \alpha_{n} \right) \, dr =0\,.
\end{equation}

Let us see if we may prove \eqref{28}. First of all, let us use a new variable given by $x:= \lambda_0/\gamma \, r$. Then,

\begin{eqnarray}\label{29}
\int_0^\infty \frac 1r \, Ai^2 \left( \frac{\lambda_0}{\gamma}\, r - \alpha_{n} \right) \, dr = \int_0^\infty \frac 1x \, Ai^2(x -\alpha_{n}) \, dx = \int_{-\alpha_{n}}^\infty \frac 1{x + \alpha_{n}}\, Ai^2(x)\, dx \nonumber \\ [2ex] = \int_{-\alpha_{n}}^0 \frac 1{x + \alpha_{n}}\, Ai^2(x)\, dx  + \int_0^\infty \frac 1{x + \alpha_{n}}\, Ai^2(x)\, dx \,.
\end{eqnarray}

The latter integral in \eqref{29} has the following upper bound:

\begin{equation}\label{30}
\int_0^\infty \frac 1{x + \alpha_{n}}\, Ai^2(x)\, dx  \le \frac 1{\alpha_{n}} \int_0^\infty Ai^2(x)\, dx \le \frac1{3^{2/3}\, \Gamma^2(1/3)\,\alpha_{n}}\,,
\end{equation}

\noindent as follows from (3.78) in \cite{VS}. Given that $\alpha_{n} \longmapsto \infty$ as $n \longmapsto \infty$ as a consequence of ii) below \eqref{12}, the integral on the lhs of \eqref{30} vanishes as $n \longmapsto \infty$. 

With respect to the former integral in the last row of \eqref{29}, the use of l'Hopital rule leads to:

\begin{equation}\label{31}
\lim_{x \to -\alpha} \frac{Ai^2(x)}{x+ \alpha_n} = 2 \lim_{x \to -\alpha} Ai(x)\, A'_i(x) = 2 Ai(-\alpha_n) Ai'(-\alpha_n)=0\,,
\end{equation}
since the numbers $-\alpha_n$ are the zeros of the Airy function $Ai(x)$. This just means that the above-mentioned integral must converge. The next step is to show that the former  integral in \eqref{29} vanishes as $n \longmapsto \infty$. First of all, note that the change $x \longmapsto -x$ transforms the integral into:

\begin{equation}\label{32}
I_n:= \frac 1{\alpha_n} \int_0^{\alpha_n} \frac{Ai^2(-x)}{1-x/\alpha_n} \, dx \,.
\end{equation}

For fixed $\alpha_n$, the Monotonic Convergence Theorem gives (note that here $x/\alpha_n$ is always positive)

\begin{equation}\label{33}
I_n = \frac 1{\alpha_n}  \sum_{k=0}^\infty \frac{1}{\alpha_n^k} \int_0^{\alpha_n} x^k\, Ai^2(-x)\, dx\,.
\end{equation}

The $k = 0$ term can be evaluated exactly, so that \eqref{33} yields 

\begin{equation}\label{34}
I_n = \frac 1{\alpha_n} \left[ \frac{1}{3^{1/3} \, \Gamma^2(1/3) } -[Ai'(-\alpha_n)]^2 \right] + O(\alpha_n^{-2})\,.
\end{equation}

The term $O(\alpha_n^{-2})$ denotes a convergent power series with respect to $\alpha^{-1}_n$, which approaches zero as $n\longmapsto \infty$, since the latter implies $\alpha_n \longmapsto \infty$. Furthermore,

\begin{equation}\label{35}
\frac{[Ai'(-\alpha_n)]^2}{\alpha_n} \approx \frac{1}{\sqrt{\alpha_n}} \longmapsto 0\,, \quad {\rm as} \quad n \longmapsto \infty\,,
\end{equation}
as a consequence of ii) below \eqref{12} and \eqref{26}, so that

\begin{equation}\label{36}
\lim_{n \to \infty} I_n =0 \,,
\end{equation}
which concludes the proof of the trace class property of the Birman-Schwinger operator.

Having shown that the Birman-Schwinger operator is trace class, we may use the zeros of the Fredholm determinant in order to evaluate of the energy of both the ground state and the first excited one.

To begin with, using the eigenfunctions $\psi_n(r)$, as defined  in \eqref{15}, we have the following split for the integral
kernel of $B_E$:

\begin{equation}\label{37}
K(r,r';E) = \omega\, \frac{1}{r^{1/2}}\, \frac{\psi_0(r)\, \psi_0(r')}{E_0-E}\, \frac{1}{{r'}^{1/2}} + M(r,r';E) = P(r,r';E) + M(r,r';E)\,,
\end{equation}

where $M(r,r';E)$ is the sum of the remainder terms. This implies that the Birman-Schwinger operator can be
written as a sum of two terms, $B_E= P+M$. 

Then, let us evaluate the zeros of the Fredholm determinant:

\begin{equation}\label{38}
0= \det[ 1-B_E] = \det[1-P-M] = \det[1-M] \det[ 1- (1-M)^{-1} P] \,.
\end{equation}

We assume that $||M||<1$. If this were the case, $\det{1-M} \ne 0$ and, therefore,

\begin{equation}\label{39}
\det[ 1- (1-M)^{-1} P] =0\,.
\end{equation}

If $A$ is a rank one operator, one has the following result for the Fredholm determinant:

\begin{equation}\label{40}
\det(1+A) = 1 + {\rm Tr}\, A \,.
\end{equation}

This is exactly the case with $P$ and, therefore, with $(1-M)^{-1} P$. Thus,

\begin{equation}\label{41}
0= \det[ 1- (1-M)^{-1} P]  = 1 - {\rm Tr}\, (1-M)^{-1} P\,.
\end{equation}

Since we assume that $||M||<1$, we have that

\begin{equation}\label{42}
(1-M)^{-1} = 1+ M +M^2 + \dots\,.
\end{equation}

Note that $M$ depends on $\omega^2$, so that in the first  order of approximation, \eqref{41} holds:

This shows that the product of the factors before the integral sign in the last term of \eqref{23} decays as rapidly as $(2n+1)^{-1}$. Then, in order to infer the convergence of the series it suffices to show that

\begin{equation}\label{28}
\lim_{n\to \infty} \int_0^\infty \frac 1r \, Ai^2 \left( \frac{\lambda_0}{\gamma}\, r - \alpha_{n} \right) \, dr =0\,.
\end{equation}

Let us see if we may prove \eqref{28}. First of all, let us use a new variable given by $x:= \lambda_0/\gamma \, r$. Then,

\begin{eqnarray}\label{29}
\int_0^\infty \frac 1r \, Ai^2 \left( \frac{\lambda_0}{\gamma}\, r - \alpha_{n} \right) \, dr = \int_0^\infty \frac 1x \, Ai^2(x -\alpha_{n}) \, dx = \int_{-\alpha_{n}}^\infty \frac 1{x + \alpha_{n}}\, Ai^2(x)\, dx \nonumber \\ [2ex] = \int_{-\alpha_{n}}^0 \frac 1{x + \alpha_{n}}\, Ai^2(x)\, dx  + \int_0^\infty \frac 1{x + \alpha_{n}}\, Ai^2(x)\, dx \,.
\end{eqnarray}

The latter integral in \eqref{29} has the following upper bound:

\begin{equation}\label{30}
\int_0^\infty \frac 1{x + \alpha_{n}}\, Ai^2(x)\, dx  \le \frac 1{\alpha_{n}} \int_0^\infty Ai^2(x)\, dx \le \frac1{3^{2/3}\, \Gamma^2(1/3)\,\alpha_{n}}\,,
\end{equation}

\noindent as follows from (3.78) in \cite{VS}. Given that $\alpha_{n} \longmapsto \infty$ as $n \longmapsto \infty$ as a consequence of ii) below \eqref{12}, the integral on the lhs of \eqref{30} vanishes as $n \longmapsto \infty$. 

With respect to the former integral in the last row of \eqref{29}, the use of l'Hopital rule leads to:

\begin{equation}\label{31}
\lim_{x \to -\alpha} \frac{Ai^2(x)}{x+ \alpha_n} = 2 \lim_{x \to -\alpha} Ai(x)\, A'_i(x) = 2 Ai(-\alpha_n) Ai'(-\alpha_n)=0\,,
\end{equation}
since the numbers $-\alpha_n$ are the zeros of the Airy function $Ai(x)$. This just means that the above-mentioned integral must converge. The next step is to show that the former  integral in \eqref{29} vanishes as $n \longmapsto \infty$. First of all, note that the change $x \longmapsto -x$ transforms the integral into:

\begin{equation}\label{32}
I_n:= \frac 1{\alpha_n} \int_0^{\alpha_n} \frac{Ai^2(-x)}{1-x/\alpha_n} \, dx \,.
\end{equation}

For fixed $\alpha_n$, the Monotonic Convergence Theorem gives (note that here $x/\alpha_n$ is always positive)

\begin{equation}\label{33}
I_n = \frac 1{\alpha_n}  \sum_{k=0}^\infty \frac{1}{\alpha_n^k} \int_0^{\alpha_n} x^k\, Ai^2(-x)\, dx\,.
\end{equation}

The $k = 0$ term can be evaluated exactly, so that \eqref{33} yields 

\begin{equation}\label{34}
I_n = \frac 1{\alpha_n} \left[ \frac{1}{3^{1/3} \, \Gamma^2(1/3) } -[Ai'(-\alpha_n)]^2 \right] + O(\alpha_n^{-2})\,.
\end{equation}

The term $O(\alpha_n^{-2})$ denotes a convergent power series with respect to $\alpha^{-1}_n$, which approaches zero as $n\longmapsto \infty$, since the latter implies $\alpha_n \longmapsto \infty$. Furthermore,

\begin{equation}\label{35}
\frac{[Ai'(-\alpha_n)]^2}{\alpha_n} \approx \frac{1}{\sqrt{\alpha_n}} \longmapsto 0\,, \quad {\rm as} \quad n \longmapsto \infty\,,
\end{equation}
as a consequence of ii) below \eqref{12} and \eqref{26}, so that

\begin{equation}\label{36}
\lim_{n \to \infty} I_n =0 \,,
\end{equation}
which concludes the proof of the trace class property of the Birman-Schwinger operator.

Having shown that the Birman-Schwinger operator is trace class, we may use the zeros of the Fredholm determinant in order to evaluate of the energy of both the ground state and the first excited one.

To begin with, using the eigenfunctions $\psi_n(r)$, as defined  in \eqref{15}, we have the following split for the integral
kernel of $B_E$:

\begin{equation}\label{37}
K(r,r';E) = \omega\, \frac{1}{r^{1/2}}\, \frac{\psi_1(r)\, \psi_1(r')}{E_1-E}\, \frac{1}{{r'}^{1/2}} + M(r,r';E) = P(r,r';E) + M(r,r';E)\,,
\end{equation}
where $M(r,r';E)$ is the sum of the remainder terms. This implies that the Birman-Schwinger operator can be
written as a sum of two terms, $B_E= P+M$. 

Then, let us evaluate the zeros of the Fredholm determinant:

\begin{equation}\label{38}
0= \det[ 1-B_E] = \det[1-P-M] = \det[1-M] \det[ 1- (1-M)^{-1} P] \,.
\end{equation}

We assume that $||M||<1$. If this were the case, $\det{1-M} \ne 0$ and, therefore,

\begin{equation}\label{39}
\det[ 1- (1-M)^{-1} P] =0\,.
\end{equation}

If $A$ is a rank one operator, one has the following result for the Fredholm determinant:

\begin{equation}\label{40}
\det(1+A) = 1 + {\rm Tr}\, A \,.
\end{equation}

This is exactly the case with $P$ and, therefore, with $(1-M)^{-1} P$. Thus,

\begin{equation}\label{41}
0= \det[ 1- (1-M)^{-1} P]  = 1 - {\rm Tr}\, (1-M)^{-1} P\,.
\end{equation}

Since we assume that $||M||<1$, we have that

\begin{equation}\label{42}
(1-M)^{-1} = 1+ M +M^2 + \dots\,.
\end{equation}

Note that $M$ depends on $\omega^2$, so that in the first  order of approximation, \eqref{41} holds:

\begin{equation}\label{43}
0=1-{\rm Tr}\, P\,.
\end{equation}

Let us write $E = E(1) = E_1+\varepsilon$. Our goal is to find an approximate expression for $\varepsilon$. As a consequence of

\begin{equation}\label{44}
P= \frac{\omega}{\varepsilon} \, |V^{1/2} \psi_1 \rangle \langle V^{1/2} \psi_1|\,,
\end{equation}
we get that (as a consquence of the fact that the functions $\psi_n$ are a basis for $L^2(\mathbb R^+)$)

\begin{eqnarray}\label{45}
{\rm Tr}\, P = \frac{\omega}{\varepsilon} \sum_{n=1}^\infty \langle \psi_n |V^{1/2} \psi_1 \rangle \langle V^{1/2} \psi_1| \psi_n\rangle =\frac{\omega}{\varepsilon}  || V^{1/2}\, \psi_1 ||_2^2 \nonumber \\ [2ex] = \frac{\omega}{\varepsilon} \left( \psi_1 , V \psi_1 \right) = \frac{\omega}{\varepsilon}  \, \frac{1}{A'_i(-\alpha_1)} \int_0^\infty \frac1r\, A_i^2 \left( \frac{\lambda_0}{\gamma}\, r - \alpha_n \right) \, dr\,.
\end{eqnarray}

Due to \eqref{43} and \eqref{45}, we conclude that

\begin{equation}\label{46}
\varepsilon = \frac{\omega}{Ai'(-\alpha_1)}\int_0^\infty \frac1r\, Ai^2 \left( \frac{\lambda_0}{\gamma}\, r - \alpha_n \right) \, dr\,,
\end{equation}
which implies that the first approximation to the ground state is given by

\begin{equation}\label{47}
E(1) = E_1 -  \frac{\omega}{Ai'(-\alpha_1)}\int_0^\infty \frac1r\, Ai^2 \left( \frac{\lambda_0}{\gamma}\, r - \alpha_n \right) \, dr
\end{equation}

Next, we make use of an important physical constraint.

\subsection{A physical bound for $||M||$}\label{subsec2}

In the above discussion, what has been missing is the existence of a crucial physical constraint. As a
matter of fact, the value of $r$ may not be arbitrarily small since the latter is just a convenient approximation.
However, due to the finite size of the particles, it is not possible to approach the origin by a distance smaller
than the particle size.  Let us assume that the particle radius is given by a constant, say $\delta$. Then, the effective
upper bound of $1/r$ has to be $1/\delta$. This has a very important consequence in order to estimate $||M||$. As we
have seen, our Birman-Schwinger operator, being trace class, is a bounded operator, so that $M$ must be bounded as well.  The integral kernel of this operator $M$ evaluated at $E=E(1)$, considering its dependence on $\omega$, is given by

\begin{eqnarray}\label{48}
M_{E(1)}(r,r')= B_{E(1)}(r,r') -P_{E(1)}(r,r') = \omega \sum_{n=2}^\infty \frac{1}{r^{1/2}}\, \frac{\psi_n(r)  \psi_n(r')}{E_n-E(1)}\, \frac{1}{{r'}^{1/2}} \nonumber \\ [2ex] = \omega \left[\sum_{n=1}^\infty \frac{1}{r^{1/2}}\, \frac{\psi_n(r) \psi_n(r')}{E_n-E(1)}\, \frac{1}{{r'}^{1/2}} - \frac{1}{r^{1/2}}\, \frac{\psi_1(r)  \psi_1(r')|}{E_1-E(1)}\, \frac{1}{{r'}^{1/2}}\right]\,.
\end{eqnarray}

Note that $E(1)=E_1-\varepsilon$. For $n \le 2$, we have that $E_n-E_1>0$, so that $E_n-E(1)= E_n- E_1+ \varepsilon > \varepsilon >0$. For $n=1$, $E_1-E(1)= \varepsilon$. 

The norm of a bounded operator $A$ on a Hilbert space is defined as (see \cite{BN,RSI}):

\begin{equation}\label{49}
||A|| = \sup_{||\psi||=1}|\langle \psi| A \psi\rangle|\,.
\end{equation}

Then, just by using \eqref{49}, we can find an upper bound for the operator norm of $M$. For any  real valued (for the sake of simplicity) $\psi$ with norm equal to one, we have:

\begin{equation}\label{50}
\left[M_{E(1)}\psi \right] (r)= \omega \sum_{n=2}^\infty \frac{1}{r^{1/2}}\, \frac{\psi_n(r) }{E_n-E(1)}\, \int_0^\infty  \frac{\psi_n(r')\psi(r')}{{r'}^{1/2}} dr'\,
\end{equation}

Thus,

\begin{eqnarray}\label{51}
\langle \psi| M_{E(1)} \psi \rangle = \omega \sum_{n=2}^\infty \frac{1}{E_n-E(1)}\, \left[\int_0^\infty  \frac{\psi_n(r)\psi(r)}{{r}^{1/2}} dr\right]^2
\end{eqnarray}

 Note that \eqref{51} is positive, since all its summands are positive, so that, it must be less or equal than

\begin{eqnarray}\label{52}
\frac{\omega}{\varepsilon} \sum_{n=2}^\infty \left[\int_0^\infty  \frac{\psi_n(r)\psi(r)}{{r}^{1/2}} dr\right]^2   \le \frac{\omega}{\varepsilon} \sum_{n=1}^\infty \left[\int_0^\infty  \frac{\psi_n(r)\psi(r)}{{r}^{1/2}} dr\right]^2
  \nonumber \\ [2ex] = \frac{\omega}{\varepsilon}  \int_0^\infty  \frac{\psi^2(r)}{r} dr \le \frac{\omega}{\varepsilon} \frac1\delta ||\psi||_2^2 = \frac{\omega}{\varepsilon}  \frac1\delta\,,\;\;\;\;
\end{eqnarray}
since the vector $\psi$ has norm equal to one.

Finally,

\begin{equation}\label{53}
||M_{E(1)}|| \le \frac{\omega}{\varepsilon}  \frac1\delta \,.
\end{equation}

Then, $||M_{E(1)}||<1$ if

\begin{equation}\label{54}
\frac{\omega}{\varepsilon} < \delta\,.
\end{equation}

\subsubsection{Another estimation.}\label{subsubsec2}

As we have seen in \eqref{14}, the eigenvalues $E_n$ are exactly opposite to $\alpha_n$, the zeros of the Airy function. Take $0 < \alpha_0 := E_2-E_1$, this means that $E_n-E_1+ \varepsilon > \alpha_0$ for all values of $n \le 2$. Thus, we may replace $\varepsilon$ by the fixed  and known constant $\alpha_0$ in the first term of \eqref{52}. If we operate exactly as in the previous paragraph, we finally have that

\begin{equation}\label{55}
||M_{E(1)}|| < \frac{\omega}{\alpha_0 \, \delta}\,,
\end{equation}

\noindent which implies that $||M_{E(1)}||<1$ if $\omega/\alpha_0 <\delta$. 

\subsection{First excited state.}\label{subsec3}

In the evaluation of the energy for the first  excited state, we have to choose as $P$ the second term of the series
\eqref{22}, which is

\begin{equation}\label{56}
P= \frac \omega\epsilon \, |V^{1/2}\, \psi_2\rangle \langle V^{1/2}\, \psi_2| \,,
\end{equation}
so that in the first  order,

\begin{equation}\label{57}
1= \frac \omega\epsilon  || V^{1/2}\, \psi_2||_2^2  \,.
\end{equation}

Consequently, in the first order of $\omega$, one has

\begin{equation}\label{58}
E(2) = E_2 - \frac{\omega}{Ai'(-\alpha_3)} \int_0^\infty \frac 1r \, Ai^2 \left( \frac{\lambda_0}{\gamma}\, r - \alpha_2 \right) \, dr
\end{equation}

In principle, the same argument may be used to evaluate other eigenvalues. Next, we may specify the eigenfunctions corresponding to these eigenvalues by using the same scheme. 

\section{On the eigenfunctions}\label{sec4}

In order to determine the eigenfunctions for $E(1)$ and $E(2)$, let us split the Birman-Schwinger operator as follows (recall that $-E=|E|$ and $\alpha_n =E_n$):

\begin{eqnarray}\label{59}
B_E = V^{1/2} (H_0+|E|)^{-1}\, V^{1/2} \nonumber \\ [2ex] = \omega \sum_{n=1}^\infty \frac{1}{|Ai'(-E_n)|^2 (E_n+|E|)} \frac{1}{r^{1/2}} \, |Ai(\frac{\lambda_0}{\gamma}\, r -E_n)\rangle \langle Ai(\frac{\lambda_0}{\gamma}\, r -E_n) |\, \frac{1}{r^{1/2}} \nonumber \\ [2ex] = \omega \sum_{n=1}^\infty \frac{1}{|Ai'(-E_{2n})|^2 (E_{2n}+|E|)} \frac{1}{r^{1/2}} \, |Ai(\frac{\lambda_0}{\gamma}\, r -E_{2n})\rangle \langle Ai(\frac{\lambda_0}{\gamma}\, r -E_{2n}) |\, \frac{1}{r^{1/2}} \nonumber \\[2ex] \oplus \omega \sum_{n=0}^\infty \frac{1}{|Ai'(-E_{2n+1})|^2 (E_{2n+1}+|E|)} \frac{1}{r^{1/2}} \, |Ai(\frac{\lambda_0}{\gamma}\, r -E_{2n+1})\rangle \langle Ai(\frac{\lambda_0}{\gamma}\, r -E_{2n+1}) |\, \frac{1}{r^{1/2}}\,.
\nonumber \\[2ex] 
\end{eqnarray}

The rank one operators in each of the infinite sums in the latter direct sum are not mutually orthogonal. Therefore, each sum should be rewritten as an infinite sum of mutually orthogonal rank one operators. It is not necessary to get into the details of the well-known orthogonalisation procedure to realise that, as we remove the component of

$$
\frac{1}{r^{1/2}}\, |Ai(\frac{\lambda_0}{\gamma}\, r -E_1)\rangle \quad {\rm from} \quad \frac{1}{r^{1/2}}\, |Ai(\frac{\lambda_0}{\gamma}\, r -E_3)\rangle\,,
$$
an additional term should be added to

$$
\frac{1}{|Ai'(-E_1)|^2 (E_1+|E|)} 
$$
in the coefficient multiplying

\begin{equation}\label{60}
\frac{1}{r^{1/2}} \, |Ai(\frac{\lambda_0}{\gamma}\, r -E_1)\rangle \langle Ai(\frac{\lambda_0}{\gamma}\, r -E_1) |\, \frac{1}{r^{1/2}}\,.
\end{equation}

Therefore, at the end of the procedure, \eqref{60} will be multiplied by an infinite series. Then, the first exact eigenvalue of the Birman-Schwinger operator is given by this infinite series multiplied by the norm $||\frac 1{r^{1/2}}\, Ai((\lambda_0/\gamma) r-E_1)||_2$, where $||-||_2$ is the norm on $L^2(\mathbb R^+)$. This series should coincide with the series provided by the Fredholm determinant.

Independently of their coefficients, the eigenfunctions for the fundamental and first excited states must be the (normalised) 
functions appearing in the projections in the first term of both orthogonal series in the last two rows in \eqref{59}.  As a consequence, the first symmetric (respectively antisymmetric) eigenfunction of the Birman-Schwinger operator is, respectively,

\begin{equation}\label{61}
\chi_1(r) =  \frac 1{r^{1/2}} \, Ai \left(\frac{\lambda_0}{\gamma}\, r -E_1 \right) \,, \qquad   \chi_2(r) =  \frac 1{r^{1/2}} Ai \left(\frac{\lambda_0}{\gamma}\, r -E_2 \right) \,.
\end{equation}

By recalling that $\chi(r) = V^{1/2} \psi(r)$, we obtain the final result for the two first normalised eigenfunctions of $H$, namely

\begin{equation}\label{62}
\Psi_1(r) = \frac{\displaystyle \, Ai \left(\frac{\lambda_0}{\gamma}\, r -E_1 \right) }{\displaystyle \left| \left| Ai \left( \frac{\lambda_0}{\gamma}\, r - E_1 \right)  \right| \right|_2 }\,=\frac {\psi_1(r)}{||\psi_1||_2}, \qquad   \Psi_2(r) = \frac{\displaystyle   Ai \left(\frac{\lambda_0}{\gamma}\, r -E_2 \right) }{\displaystyle \left| \left| Ai \left( \frac{\lambda_0}{\gamma}\, r - E_2 \right) \right| \right|_2 }\,=\frac {\psi_2(r)}{||\psi_2||_2}.
\end{equation} 

In either case, these are the normalised eigenfunctions at the zeroth order. As was stated before, both series in \eqref{59} are not orthogonal sums, so that in order to obtain the eigenfunctions to higher order, we may use the Gram-Schmidt orthonormalisation method. Let us briefly recall this method here. In fact, we start with a countable set of linearly independent vectors, $[y_1,y_2,\dots,y_n,\dots]$ in an inner product space $X$ and we want to obtain a sequence of orthonormal vectors $[x_1,x_2, \dots,x_n,\dots]$ spanning the same subspace. The procedure is well known and is defined by the following inductive prescription:

\begin{equation}\label{64}
x_1= \frac{y_1}{||y_1||}\,, \qquad x_n:= \frac{\omega_n}{||\omega_n||}\,, \qquad \omega_n= y_n -\sum_{i=1}^{n-1} \langle x_i|y_n\rangle x_i\,,
\end{equation}

\noindent so that for $n=2$ in the Gram-Schmidt procedure we have that 
$$\omega_2= y_2 - \left( x_1,y_2\right) x_1$$
 
Here we wish to obtain the first order expression for the eigenfunctions pertaining to $E(1)$ and $E(2)$ respectively. By restricting ourselves to $E(1)$, we have

\begin{equation}\label{65}
x_1 \equiv \frac{\chi_1(r)}{||\chi_1(r)||_2}\,, \qquad  y_3= Ai(\frac{\lambda_0}{\gamma}\, r -E_{3}) \, \frac{1}{r^{1/2}}\,
\end{equation}

\noindent so that

\begin{equation}\label{66}
\left( x_1,y_3\right) =  \int_0^\infty Ai \left(\frac{\lambda_0}{\gamma}\, r -E_1 \right) \, Ai \left(\frac{\lambda_0}{\gamma}\, r -E_3 \right) \, \frac{dr}{r}\,,
\end{equation}

\noindent which implies that

\begin{eqnarray}\label{67}
\omega_3(r)= Ai(\frac{\lambda_0}{\gamma}\, r -E_{3}) \, \frac{1}{r^{1/2}}-\left( x_1,y_3\right) \,  x_1 = Ai(\frac{\lambda_0}{\gamma}\, r -E_{3}) \, \frac{1}{r^{1/2}}\, \nonumber \\ [2ex] -\, \frac {\int_0^\infty Ai \left(\frac{\lambda_0}{\gamma}\, r -E_1 \right) \, Ai \left(\frac{\lambda_0}{\gamma}\, r -E_3 \right) \, \frac{dr}{r}}{\int_0^\infty Ai^2 \left(\frac{\lambda_0}{\gamma}\, r -E_1 \right) \,  \frac{dr}{r}}\, Ai(\frac{\lambda_0}{\gamma}\, r -E_{1}) \, \frac{1}{r^{1/2}}
\end{eqnarray}

It might be worth noting that \eqref{67} can be recast as 

\begin{equation}\label{68}
 \omega_3(r)= \chi_3(r)\, -  \frac {\int_0^\infty \chi_1(r)\, \chi_3(r)\, \, dr}{\int_0^\infty \,  \chi^2_1(r) dr}\, \chi_1(r)=\chi_3(r)\, -  \frac {\left( \chi_1, \chi_3\right)}{ ||\chi_1||^2_2}\, \chi_1(r)
\end{equation}

As an immediate consequence of the fact that $\chi(r) = V^{1/2} \psi(r)$, the corresponding eigenfuction (to be normalised) for our Hamiltonian reads:

\begin{eqnarray}\label{69}
r^{1/2}\,\omega_3(r)=  Ai(\frac{\lambda_0}{\gamma}\, r -E_{3})\, - \nonumber \\ [2ex] \frac {\int_0^\infty Ai \left(\frac{\lambda_0}{\gamma}\, r -E_1 \right) \, Ai \left(\frac{\lambda_0}{\gamma}\, r -E_3 \right) \, \frac{dr}{r}}{\int_0^\infty Ai^2 \left(\frac{\lambda_0}{\gamma}\, r -E_1 \right) \,  \frac{dr}{r}}\, Ai(\frac{\lambda_0}{\gamma}\, r -E_{1}) \,= \nonumber \\ [2ex] \psi_3(r)\, -  \frac {\left( \psi_1, \frac{1}{r}\,\psi_3\right)}{ \left( \psi_1, \frac{1}{r}\,\psi_1\right)}\, \psi_1(r)
\end{eqnarray}

Of course, the latter function is not orthogonal to $\psi_1$ since:

\begin{eqnarray}\label{70}
\left(\psi_1, r^{1/2}\,\omega_3 \right)=\left(\psi_1, \psi_3 \right)  -  \frac {\left( \psi_1, \frac{1}{r}\,\psi_3\right)}{ \left( \psi_1, \frac{1}{r}\,\psi_1\right)}\,  ||\psi_1  ||^2_2= \nonumber \\ [2ex] -  \frac {\left( \psi_1, \frac{1}{r}\,\psi_3\right)}{ \left( \psi_1, \frac{1}{r}\,\psi_1\right)}\,  ||\psi_1  ||^2_2\, \neq 0
\end{eqnarray}

\smallskip

Similarly, for the two lowest eigenfunctions in the series appearing in the second summand of the direct sum in \eqref{59} we have:

\begin{eqnarray}\label{71}
\omega_4(r)= Ai(\frac{\lambda_0}{\gamma}\, r -E_{4}) \, \frac{1}{r^{1/2}}-\left( x_2,y_4\right) \,  x_2 = Ai(\frac{\lambda_0}{\gamma}\, r -E_{4}) \, \frac{1}{r^{1/2}}\, \nonumber \\ [2ex] -\, \frac {\int_0^\infty Ai \left(\frac{\lambda_0}{\gamma}\, r -E_2 \right) \, Ai \left(\frac{\lambda_0}{\gamma}\, r -E_4 \right) \, \frac{dr}{r}}{\int_0^\infty Ai^2 \left(\frac{\lambda_0}{\gamma}\, r -E_2 \right) \,  \frac{dr}{r}}\, Ai(\frac{\lambda_0}{\gamma}\, r -E_2) \, \frac{1}{r^{1/2}},
\end{eqnarray}

\smallskip

\noindent which can be rewritten as

\begin{equation}\label{72}
 \omega_4(r)= \chi_4(r)\, -  \frac {\int_0^\infty \chi_2(r)\, \chi_4(r)\, \, dr}{\int_0^\infty \,  \chi^2_2(r) dr}\, \chi_2(r)=\chi_4(r)\, -  \frac {\left( \chi_2, \chi_4\right)}{ ||\chi_2||^2_2}\, \chi_2(r)
\end{equation}

\smallskip

Therefore,

\begin{eqnarray}\label{73}
r^{1/2}\,\omega_4(r)=  Ai(\frac{\lambda_0}{\gamma}\, r -E_4)\, - \nonumber \\ [2ex] \frac {\int_0^\infty Ai \left(\frac{\lambda_0}{\gamma}\, r -E_2 \right) \, Ai \left(\frac{\lambda_0}{\gamma}\, r -E_4 \right) \, \frac{dr}{r}}{\int_0^\infty Ai^2 \left(\frac{\lambda_0}{\gamma}\, r -E_2 \right) \,  \frac{dr}{r}}\, Ai(\frac{\lambda_0}{\gamma}\, r -E_2) \,= \nonumber \\ [2ex] \psi_4(r)\, -  \frac {\left( \psi_2, \frac{1}{r}\,\psi_4\right)}{ \left( \psi_2, \frac{1}{r}\,\psi_2\right)}\, \psi_2(r)
\end{eqnarray}

\section{Concluding Remarks}

In this note we have developed an application of the  Birman-Schwinger operator method to the case of the radial component of the Cornell potential. The use of the formalism appears to be simpler than other approaches, such as purely numerical or semianalytical  methods, among which we wish to mention the Nikiforov-Uvarov formalism. The exact solutions of the radial components of the potential, after treating it by means of the Birman-Schwinger operator method,  are expressed in terms of Airy functions. Thus, the possibility to study analytically relevant features of QCD in the non-perturbative regime may be facilitated. 
The alternative to this approach would be the diagonalisation of the kinetic energy term of the Hamiltonian in a basis chosen in a way of getting radial wave functions with nodes determined by the size of the system to be reproduced under the conditions of confinement. The calculation would then proceed by diagonalising the remaining components of the Hamiltonian in the space of eigenstates of the kinetic energy term. Other possibilities, as mentioned in the introduction of this paper, would then proceed in the same manner by using effective Hamiltonians parametrised in such a manner to reproduce observed features like masses, electromagnetic moments, etc. Without exceptions these other ways to proceed are similarly involved. We think that our approach, from the mathematical point of view, is better justified. Work is in progress about further applications of the formalism to the low-energy meson spectra.

\section*{Acknowledgements}

We wish to express our gratitude to Profs. Mervin Lawrence Glasser and Luis Miguel Nieto for fruitful discussions and suggestions that have helped much in the present research. Financial support is also acknowledged to the Spanish MCIN with funding from the European Union Next Generation EU, PRTRC17.11, and the Consejer\'ia de Educaci\'on from the JCyL through the QCAYLE project, as well as MCIN projects PID2020-113406GB-I00 and RED2022-134301-T. This work is part of the PIP 2081 of the CONICET, Argentina.

\end{document}